# Nanoscale Magnetic Imaging using Circularly Polarized High-Harmonic Radiation

Ofer Kfir[1,2*], Sergey Zayko[1*], Christina Nolte[3], Murat Sivis[1], Marcel Möller[1], Birgit Hebler[4], Sri Sai Phani Kanth Arekapudi[4], Daniel Steil[3], Sascha Schäfer[1], Manfred Albrecht[4], Oren Cohen[2], Stefan Mathias[3], and Claus Ropers[1,5]

[1]University of Göttingen, 4th Physical Institute, Göttingen 37077, Germany.
[2]Solid State Institute and Physics Department, Technion – Israel Institute of Technology, Haifa 32000, Israel.
[3]University of Göttingen, 1st Physical Institute, Göttingen 37077, Germany.
[4]Institute of Physics, University of Augsburg, Augsburg 86159, Germany.
[5]International Center for Advanced Studies of Energy Conversion (ICASEC), University of Göttingen, Germany.
Email for correspondence: ofer.kfir@phys.uni-goettingen.de, claus.ropers@uni-goettingen.de
*These authors contributed equally to this work

**This work demonstrates nanoscale magnetic imaging using bright circularly polarized high-harmonic radiation. We utilize the magneto-optical contrast of worm-like magnetic domains in a Co/Pd multilayer structure, obtaining quantitative amplitude and phase maps by lensless imaging. A diffraction-limited spatial resolution of 49 nm is achieved with iterative phase reconstruction enhanced by a holographic mask. Harnessing the unique coherence of high harmonics, this approach will facilitate quantitative, element-specific and spatially-resolved studies of ultrafast magnetization dynamics, advancing both fundamental and applied aspects of nanoscale magnetism.**

Nanoscale magnetic structures exhibit a rich variety of patterns, textures, and topological states[1,2], governed by the complex interplay of multiple spin-coupling mechanisms[3]. Mapping such spin configurations and their dynamic response is essential for the development of functional nanomagnetic systems[4–10]. Magnetic imaging with high spatial and ultrafast temporal resolution is possible using circularly polarized extreme ultraviolet (EUV) and soft-X-ray radiation. However, short-wavelength magneto-optical microscopy[7,10] is presently limited to synchrotrons and free-electron lasers (FELs), although recent developments in polarization-controlled high-harmonic generation promise a laboratory-scale implementation[11,12].

High-order harmonic generation (HHG)[13] is a process that gives rise to spatially and temporally coherent EUV[14] and soft-X-ray beams with unique features. The spectral bandwidth of high-harmonic radiation supports attosecond pulses[15,16] and can be used to probe multiple chemical elements[17,18] or to create stable EUV frequency combs[19]. Imaging experiments using high-harmonic radiation[20–23] benefit from the source's coherence, which enables high spatial resolution at a large field-of-view[24,25]. High-harmonic imaging has consistently proven capable of nanoscale resolution for high-contrast, lithographically produced objects. On the other hand, the mapping of chemical and electronic contrast, polarization anisotropies[26], chiral features, or spin textures has remained challenging, despite a great potential for applications. In particular, the spectral region accessible by high harmonics spans the magneto-optical activity range of widely-used ferromagnetic materials, facilitating, for example, spectroscopic[17] or diffractive[27] probing of ultrafast magnetism.

Generally, magneto-optical imaging with EUV and X-ray radiation combines element specificity and an *in-situ* compatibility with currents or strong electric and magnetic fields[28], which has allowed for dynamical studies of domain walls[29], magnetic vortices[7] and skyrmions[10]. Full-field magneto-optical imaging was pioneered by Eisebitt *et al.*[30], employing Fourier transform holography (FTH). In this scheme, X-ray magnetic circular dichroism (XMCD) provides phase and amplitude contrast of the magnetization component parallel to the circularly polarized X-ray beam. As the magneto-optical contrast[31] is typically weak and suffers from non-magnetic absorption, to date, XMCD-based microscopy is available exclusively at large-scale EUV and X-ray sources. The proliferation of such schemes based on laser-driven, table-top implementations requires both high flux and circular polarization, two traditional challenges for HHG.

In this work, we use circularly polarized high-harmonic radiation to reconstruct nanoscale magnetic structures by FTH and iterative phase retrieval for coherent diffractive imaging (CDI). Specifically, we map worm-like magnetic domains in a cobalt/palladium (Co/Pd) multilayer stack using XMCD contrast within the cobalt M-edge spectral region (59 eV). We measure magneto-optical absorption and phase shifts with a spatial resolution down to 49 nm. The exceptional coherence of HHG allows us to enhance the magnetic signal using intense reference waves from tailored holographic masks.

The experimental scheme is shown in Fig. 1: A pulsed laser beam from an amplified Ti:Sapphire laser system (pulse duration 45 fs, repetition rate 1 kHz, central wavelength 800 nm, pulse energy 2 mJ) is converted by an in-line "MAZEL-TOV" apparatus[32] to a beam with superimposed circularly-polarized fundamental and second-harmonic fields of opposite helicities. This bi-chromatic laser field generates circularly polarized high harmonics in a He-filled gas cell (8 mm diameter, pressure of 500 mbar). The generation phase-matching conditions[33] are optimized by tuning the gas pressure, the position of the focus, and the laser confocal parameter. A 150-nm-thick Al foil blocks the bi-chromatic beam.

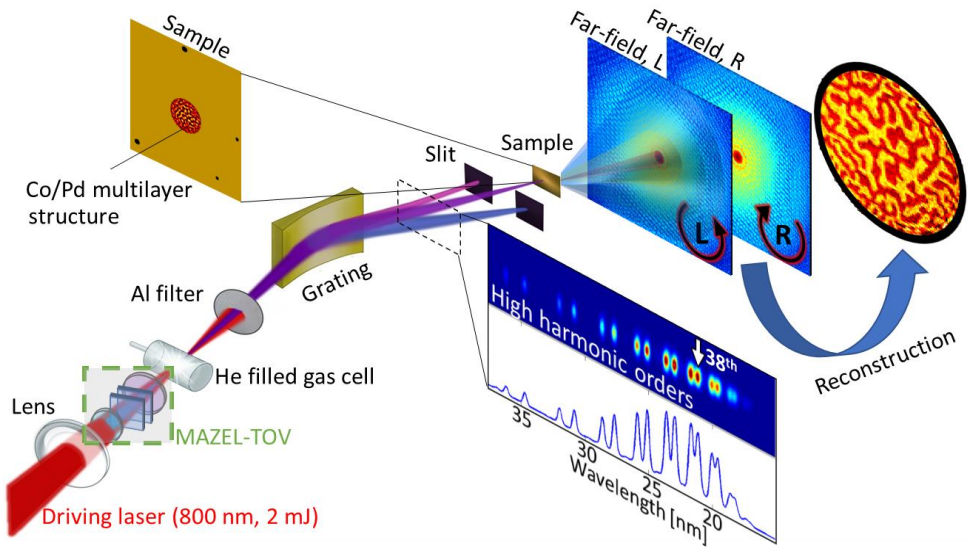

Figure 1. A scheme of the experimental setup. A bi-chromatic circularly polarized beam tailored by a MAZEL-TOV apparatus generates circularly polarized high-order harmonics in a He-filled gas cell. The harmonics are spatially dispersed and refocused onto the sample by a single toroidal grating, and the 38th harmonic (marked on the spectrum) is selected by a slit. The scattering pattern from the sample is recorded with a CCD camera for left-hand (L) and right-hand (R) circular polarization of the high-harmonic beam. The magnetic domain pattern is retrieved from the two diffraction patterns by Fourier transform holography and iterative phase retrieval. The sample (sketch, top left) shows the four reference holes and the worm-like domain structure on the central aperture.

The imaging system comprises a toroidal diffraction grating, a slit, the sample, and a charge-coupled device (CCD) camera. The toroidal grating spatially disperses the harmonics and re-focuses the selected 38th harmonic (21 nm wavelength, 59 eV photon energy) onto the sample. We optimize the rotation angle of the quarter-wave plate in the MAZEL-TOV device to achieve the correct selection rule for circularly polarized high-harmonic generation, which is evident from the suppression of every third harmonic order[11,12,34,35]. The sample is a 200-nm-thick Si membrane, prepared with a magnetic multilayer Co/Pd stack on the front side. The magnetic film comprises 9 pairs of cobalt and palladium layers, $[Co_{(0.47nm)}/Pd_{(0.75nm)}]_9$, which results in worm-like magnetic domains with out-of-plane anisotropy after demagnetization (alternating external out-of-plane magnetic field of decreasing strength, see Appendix). The backside of the substrate is covered with a 180-nm gold layer. Using focused ion beam (FIB) etching, the EUV-opaque gold film is removed to form a central aperture. To complete the holographic mask, four reference holes with varying diameters are milled through the entire structure (see SEM micrograph in Fig. 2c).

Figure 2 displays the scattering data and reconstructions of the magnetic domain structure. The diffraction pattern recorded by illuminating the sample with a left-hand circularly polarized HHG beam is shown in Fig. 2a. For each reference hole, the Fourier transform of the diffraction pattern yields a holographic reconstruction of the complex wave exiting the central aperture (cf. Fig. 2b, see also Appendix). Individual reconstructions of the exit waves $f_{obj,L}(\vec{r})$ or $f_{obj,R}(\vec{r})$ for left-hand (L) or right-hand (R) circularly polarized illumination, respectively, exhibit a small magneto-optical signal on a large non-magnetic background. Forming the ratio of the exit waves, we directly access the XMCD phase and absorption in a quantitative manner. The measured phase difference $\phi$ and amplitude ratio $\rho$ are related to the magnetization component parallel to the beam, $M_z(\vec{r})$, via

$$\frac{f_{obj,L}(\vec{r})}{f_{obj,R}(\vec{r})} = \exp\left[2ikd(\Delta\delta + i\Delta\beta)\left(\frac{M_z(\vec{r})}{|\vec{M}|}\right)\right] =: \rho e^{i\phi} \qquad (1)$$

(see details in the Appendix). Here, $\Delta\delta, \Delta\beta$ are the dichroic refraction and absorption coefficients of cobalt[31], respectively, $k$ is the wavenumber, $d$ is the overall thickness of the magneto-optically active material (Co), and $|\vec{M}|$ is its saturation magnetization. Figures 2d and 2e show the XMCD phase-contrast images obtained by FTH and CDI reconstruction, respectively. At the probing wavelength, the phase images provide somewhat higher contrast than the amplitude images and are less sensitive to noise, as the dichroic magneto-optical refraction is larger than the absorption.

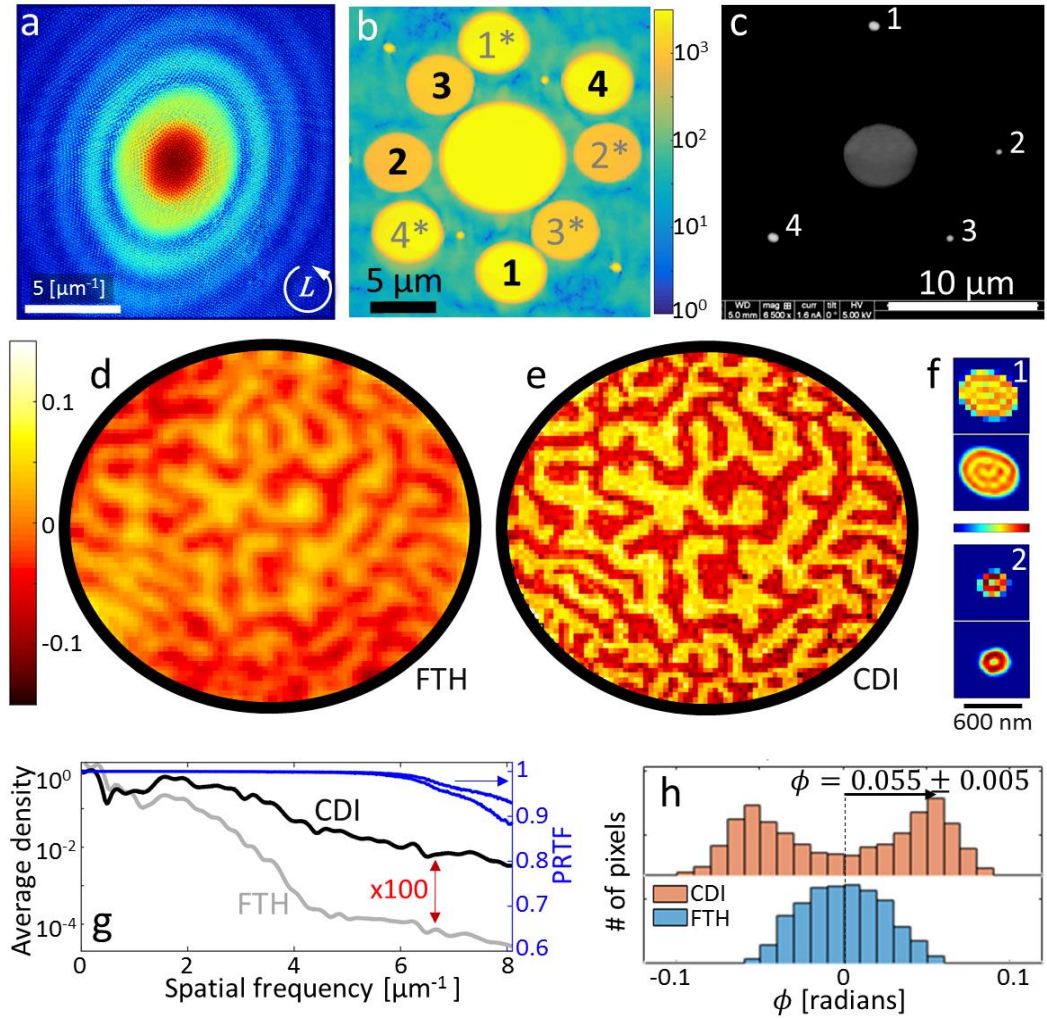

Figure 2. Experimental results and the retrieval of nanoscale magnetic domain structure using high-order harmonics. (a) Diffraction pattern recorded with a left-hand circularly polarized 21 nm high-harmonic beam (logarithmic color scale, 341 seconds total integration time, $9 \cdot 10^8$ detected photons, corrected for spherical aberration by projection onto the Ewald sphere). (b) Fourier transform magnitude of the recorded diffraction pattern (logarithmic color scale). The holographic reconstructions from reference holes are marked 1 to 4 (conjugate holograms: 1* to 4*). (c) Scanning electron micrograph (SEM) of the gold-coated side of the sample, showing the central aperture (gray, 5 μm diameter) and four reference holes (white, numbered). (d,e) Magneto-optical phase contrast images of the worm-like domain structure, obtained by FTH (reference hole 2, 250 nm diameter) and by CDI reconstruction, respectively (true pixel resolution, no interpolation). Positive (negative) phase indicates magnetization parallel (antiparallel) to the HHG beam. (f) Magnitude (linear scale, magnified) of the reconstructed exit-waves of reference holes 1 and 2, at true-pixel resolution (top) and interpolated (bottom), illustrating the presence of high-order waveguide modes. (g) Left axis: Fourier spectral density of phase contrast images (azimuthally averaged). Right axis: Phase retrieval transfer functions (PRTF) for both incident polarizations, demonstrating high fidelity of the CDI reconstructions. (h) Histogram of the measured phase, as reconstructed using CDI and FTH, allows for quantitative evaluation of the magneto-optical phase.

To retrieve the exit-wave of our sample by CDI, we employed an iterative error reduction algorithm[36] incorporating the FTH reconstruction for the real-space support (details are in the Appendix). The CDI reconstruction (Fig. 2e) yields a high-resolution refinement of the hologram, with sharp domain boundaries as well as a flat phase and amplitude within the domains. In addition, the reconstructed exit-wave of the reference holes (Fig. 2f) includes multiple modes, with modulation at the level of a single pixel (60 nm in this case) corresponding to scattering that reaches the edges of the CCD detector. Thus, the diffracted reference field interferes with the object wave over the entire detector. An analysis of the azimuthally averaged spatial frequencies (Fig. 2g) of the reconstructions from FTH and CDI shows that the CDI phase map exhibits high-frequency components, which are damped in FTH by two orders of magnitude. Moreover, a broad peak around 1.7 µm$^{-1}$ indicates a typical domain size of 300 nm. The phase retrieval transfer functions (PRTF[37], blue curves in Fig. 2g) demonstrate a reliable reconstruction throughout the detected diffraction pattern. The flat dichroic contrast within the reconstructed domains shows that the magnetization is locally saturated out-of-plane and oriented either parallel (bright contrast) or antiparallel (dark contrast) to the illuminating beam. This allows for a quantitative measurement of the dichroic phase, $\phi$, and the magneto-optical phase coefficient $\Delta\delta$. From the histogram of measured phases (cf. Fig. 2h), we determine a dichroic phase shift of $\phi = 0.055 \pm 0.005$, corresponding to $\Delta\delta = 0.022 \pm 0.002$ for the given film thickness (see Appendix for details). This value is larger than that obtained for free-standing cobalt[31], but consistent with measurements in multilayer stacks[38].

Based on this demonstration of diffraction-limited magnetic imaging, the spatial resolution can be further improved by detecting larger scattering angles. In Figure 3, we present reconstructions of a different magnetic domain structure at a single-pixel, 49-nm resolution (object aperture: 4 µm, otherwise similar mask geometry). The recorded diffraction pattern (Fig. 3a) covers spatial frequencies up to 10.2 µm$^{-1}$ (corner: 14.3 µm$^{-1}$) with high visibility across the CCD (see zoom of the top-right corner). The images of the dichroic phase and amplitude (Figs. 3b,c) rival the quality and resolution of state-of-the-art experiments at FELs and synchrotrons[39,40], at an even larger field-of-view. The reconstructed images clearly resolve curves, edges, and other fine features of the domains. Again, diffraction-limited resolution is achieved, as evident from the step-like contrast changes between adjacent domains (lineout in Fig. 3d).

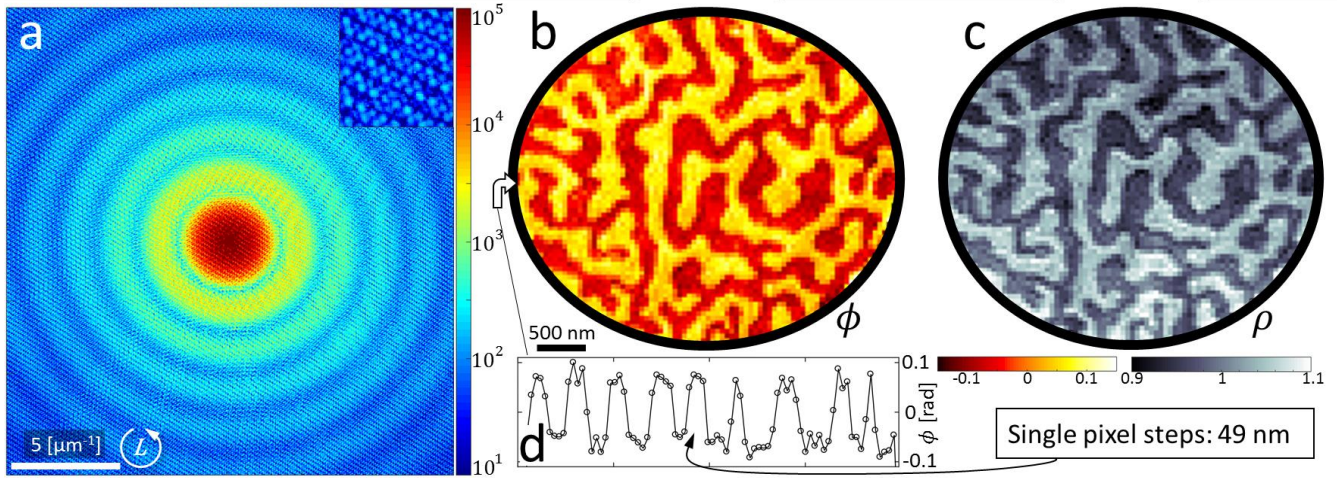

Figure 3. Imaging of magnetic domains with 49-nm resolution. (a) Recorded diffraction pattern with high speckle visibility (see zoomed corner). Prominent diffraction rings arise from the four reference holes in the holographic mask (circular shape, diameter of ~550 nm). (b,c) CDI reconstructions of the magnetic domain structure with (b) phase and (c) amplitude magneto-optical contrast, respectively.

Our analysis illustrates that applying CDI to a data set with holographic information is advantageous on multiple levels[41,42], especially for weakly scattering objects, as in XMCD. First, the hologram assists a rapid convergence of the iterative algorithm, with additional redundancy provided by multiple reference holes of varying sizes. This results in an unambiguous CDI reconstruction directly validated by FTH[21]. Second, the hologram provides for a data-based support of the exit wave, accurate to the size of the smallest reference hole. This may be particularly useful for structures with a complicated support. Third, the signal from a weakly scattering object is enhanced by the interference with a strong reference wave, which enables a CDI reconstruction. Finally, while FTH resolution suffers for larger reference holes, the CDI algorithm is fully capable of reconstructing their complex exit-wave patterns, leading to a diffraction-limited resolution. Specifically, guided modes with high spatial frequencies suffer less attenuation upon propagation in wider reference holes[22], and thus, can enhance the high-resolution information at large spatial frequencies. In these experiments, the scattering from large reference holes provides substantial diffraction intensity across the detector (see rings in Figs. 2a,3a), which is instrumental for reconstructing the weak magneto-optical scattering with diffraction-limited resolution.

In conclusion, this work reports the first nanoscale magnetic imaging with high-order harmonic radiation. We use circularly-polarized illumination to map a randomly-formed magnetic domain pattern, and reach a spatial resolution of 49 nm by coherent diffractive imaging. Our experiment shows that for weakly scattering objects, structured reference waves contribute to the robustness, stability and resolution of CDI phase retrieval. This approach can be further developed to achieve resolutions of 10 nm and less, for example, by employing keV-scale high harmonics[43–45] to access L-edge XMCD contrast. The discrete harmonic spectrum may also allow for an extension to hyperspectral imaging[40,46,47], facilitating multiple-element contrast and the investigation of spatio-temporal dynamics in magnetic heterostructures or skyrmion systems. Besides applications in magnetism, we believe that table-top nano-imaging based on circularly polarized high-harmonic radiation will contribute to the fundamental understanding of microscopic chiral phenomena.

# Appendix

**Sample preparation.** The multilayer stack, $Pd_{(2\,nm)}/[Co_{(0.47\,nm)}/Pd_{(0.75\,nm)}]_9/Pd_{(2\,nm)}/Cr_{(1.5\,nm)}$, was deposited via DC magnetron sputtering at room temperature at a 4.5 mTorr Ar partial pressure. The thicknesses of the individual materials were pre-calibrated by X-ray reflectometry (XRR). Here, Cr acts as an adhesion layer, followed by a Pd layer that promotes a (111) texture of the Co/Pd-multilayers. A Pd cap layer provides an oxidation barrier. The substrate is the front side of a 200-nm thick Si membrane, whereas the central aperture and the reference holes are milled from the Au-coated backside (180 nm). A typical worm-like domain pattern[48] is formed by saturating the out-of-plane magnetization in an external field, followed by a set of field polarity inversions, reducing the field magnitude by 15% in each inversion.

**Data acquisition and analysis.**

For the data in Fig. 2, two individual diffraction patterns (exposure times of 330 sec and 11 sec) were recorded with a CCD camera (1340x1300 pixels, 20 µm pixel size) for each helicity. Combining the diffraction patterns of these two exposure times increases the total dynamic range of the diffraction pattern. After a dark image subtraction, the left-hand and right-hand circularly polarized diffraction patterns are centered and mapped onto the Ewald sphere to account for spherical aberrations induced by the flat CCD detector. The real-space support for the iterative phase retrieval process is derived by thresholding the inverse Fourier transformation of the measured far-field intensities (Fig. 2b), and subsequent deconvolution of the object's support from its autocorrelation. A significant improvement of the image quality is achieved already after a few hundred iterations of the error reduction algorithm[36]. A consistent convergence is evident from the PRTF, which is an azimuthal average over the mean phase, $\langle \exp[i\psi]\rangle$, of 20 individual reconstructions initiated with a random far-field phase. The data in Fig. 2e uses $1.8*10^5$ iterations.

The data in Fig. 3 was acquired under similar conditions. The scattering pattern was acquired for 600 seconds per helicity, and the overexposed central part was replaced with the data from diffraction patterns composed of 30 accumulations with 6 sec exposure time each. The central aperture in the holographic mask has a 4 µm diameter. The reference holes are equidistantly displaced 6.6 µm away from the center of the aperture and have circular shapes with diameters of ~550 nm.

**Holographic reconstruction.** The complex wave exiting the sample can be written as a superposition of those from the central aperture and the reference holes,

$$f_{exit}(\vec{r}) = f_{obj}(\vec{r}) + \sum_l f_{ref,l}(\vec{r} - \vec{r}_l).$$

Here, $\vec{r}$ is the two-dimensional radius vector in the sample plane, $f_{obj}(\vec{r})$ is the exit wave of the central aperture, and $l = 1,2,3,4$ is an index denoting the reference holes, where $\vec{r}_l$ and $f_{ref,l}(\vec{r} - \vec{r}_l)$ indicate the position and the exit wave of reference hole $l$, respectively. In the far-field (Fraunhofer approximation[49]), the intensity distribution measured by the CCD is given by

$$I_{CCD} \propto \left|FT[f_{obj}(\vec{r})]\right|^2 + \left|\sum_l FT[f_{ref,l}(\vec{r} - \vec{r}_l)]\right|^2 + 2\sum_l Re\left[FT[f_{obj}(\vec{r})]\left(FT[f_{ref,l}(\vec{r} - \vec{r}_l)]\right)^*\right],$$

where FT is the two-dimensional Fourier transform. Thus, an inverse Fourier transform of this intensity pattern results in multiple reconstructions of the object's exit wave, one for every reference hole:

$$f_{rec,l} = f_{obj}(\vec{r}) \otimes f_{ref,l}^*(-\vec{r} - \vec{r}_l) \approx f_{obj}(\vec{r} + \vec{r}_l),$$

where $\otimes$ denotes a convolution, and the final approximation corresponds to a point-like reference hole. Thus, the reconstruction of hole $l$ is placed on the opposite side of its physical position on the sample. A conjugated image for hole $l$ is given by $f_{conj,l} = f_{obj}^*(-\vec{r}) \otimes f_{ref,l}(\vec{r} - \vec{r}_l) \approx f_{obj}^*(-\vec{r} + \vec{r}_l)$.

**Quantitative estimation of the magneto-optical coefficient, $\Delta\delta$.** The magneto-optical activity of cobalt is expressed in the refractive index[31], $n_\pm = (1 - (\delta \pm \Delta\delta)) - i(\beta \pm \Delta\beta)$. The magneto-optical refraction coefficients are proportional to the magnetization projected onto the optical axis[50], $M_z/|\vec{M}|$, so that for left-hand circular polarization, parallel/antiparallel magnetization acquires a +/- sign. Thus, a plane wave transmitted through a magnetic sample results in the exit wave $f_{exit, {L \atop R}} = e^{-ikdn_\pm} = e^{-kd(\beta\pm\Delta\beta)}e^{-ikd(1-(\delta\pm\Delta\delta))}$. Eq. (1) is retrieved by dividing the exit waves for left-hand (L) and right-hand (R) circular polarizations, and accounting for the magnetization strength and orientation, $M_z/|\vec{M}|$. For locally saturated magnetization parallel or antiparallel to the beam, the magneto-optical refraction is $\Delta\delta = \phi/2kd$, where $\phi$ is the measured dichroic phase shift. From the peaks of the phase contrast histogram (Fig. 2h), the dichroic phase shift is $\phi = 0.055 \pm 0.005$. Considering the total thickness of the cobalt layers, $d_{Co}$=4.23 nm ($kd_{Co}$=1.26), the magneto-optical refraction is $\Delta\delta = 0.022 \pm 0.002$. Our estimation for $\Delta\delta$ assumes that the illumination is purely circularly polarized, and that the magnetization in the fully saturated regions is exactly parallel/antiparallel to the beam.

**Notations for light helicity and magneto-optical interaction.** For polarization helicities, we use a notation[51] in which the electric field of left-handed circular polarization is $\vec{E} \propto [\hat{x}\cos(\omega t - kz) - \hat{y}\sin(\omega t - kz)]$. $\hat{x}$ and $\hat{y}$ are the cartesian unit-vectors in the polarization plane, $k$ is the wave number, $\omega$ is the angular frequency, $t$ is time, and $z$ is the position along the optical axis. For a given spatial wavefunction, $f(\vec{r})$, e.g., at the exit-plane of the sample, the physical electric field can be written in complex form as $\vec{E}_{L,R}(\vec{r}, t) = Re[E_0 \hat{e}_{L,R} e^{i\omega t} f(\vec{r})]$. Here, $\hat{r}$ is the 2-dimensional radius vector in the polarization plane. $E_0$ is the field's complex amplitude (including a constant phase), and $Re[\,]$ denotes the real part. $L, R$ mark left-hand and right-hand circular polarization, with $\hat{e}_{L,R}$ as the normalized polarization vector, $\hat{e}_L = \frac{1}{\sqrt{2}}(\hat{x} + i\hat{y})$, $\hat{e}_R = \frac{1}{\sqrt{2}}(\hat{x} - i\hat{y})$.

As for the notation for the magneto-optical interaction, the term "parallel" magnetization is defined here differently than in the work by Valencia et al.,[31]. Note that we describe the magnetization as being parallel or antiparallel to the beam's propagation direction. In Ref. [31], "parallel" refers to the relation between the magnetization and the respective polarization helicity of the beam. These terminologies coincide for left-hand polarization.

**Acknowledgements**

This work was funded by the Deutsche Forschungsgemeinschaft (DFG) within SFB 755 "Nanoscale photonic imaging" (Project C08). We gratefully acknowledge discussions and support from Tim Salditt, Christoph Lienau and Heiko Kollmann.